%% file: main.tex
\documentclass[reprint,aip,pop,superscriptaddress]{revtex4-1}
\usepackage{amsmath}
\usepackage{amssymb}
\usepackage{cancel}
\usepackage{stmaryrd}
\usepackage{amsfonts}
\usepackage[T1]{fontenc}
\usepackage{bm}
\usepackage{color}
\usepackage{graphicx}
\usepackage{comment}
\usepackage{dcolumn}
\usepackage{lipsum}
\includecomment{comment}  

\DeclareSymbolFont{operators}{OT1}{cmr}{m}{n}
\DeclareSymbolFont{letters}{OML}{cmm}{m}{it}
\DeclareSymbolFont{symbols}{OMS}{cmsy}{m}{n}
\DeclareSymbolFont{largesymbols}{OMX}{cmex}{m}{n}

\usepackage{hyperref}
\hypersetup{
    unicode=false,  
    pdftoolbar=true,
    pdfmenubar=true,
    pdffitwindow=false,
    pdfstartview={FitH},
    pdftitle={Closure problem},
    pdfauthor={Eero Hirvijoki},
    pdfsubject={Runaway electrons},
    pdfcreator={Eero Hirvijoki},
    pdfproducer={GNU Make},
    pdfkeywords={Runaway electrons} {Closure} {one-fluid equations},
    pdfnewwindow=true,
    colorlinks=true,
    linkcolor=blue,
    citecolor=blue,
    filecolor=blue,
    urlcolor=blue
}

\begin{document}

\preprint{AIP/123-QED}

\title{A fluid-kinetic framework for self-consistent runaway-electron simulations}

\author{Eero Hirvijoki}
\affiliation{Princeton Plasma Physics Laboratory, Princeton, New Jersey 08543, USA}
\email{ehirvijo@pppl.gov}
\author{Chang Liu}
\affiliation{Princeton Plasma Physics Laboratory, Princeton, New Jersey 08543, USA}
\author{Guannan Zhang}
\affiliation{Oak Ridge National Laboratory Oak Ridge, Tennessee 37831-8071, USA}
\author{Diego del-Castillo-Negrete}
\affiliation{Oak Ridge National Laboratory Oak Ridge, Tennessee 37831-8071, USA}
\author{Dylan Brennan}
\affiliation{Princeton Plasma Physics Laboratory, Princeton, New Jersey 08543, USA}


\date{\today}

\begin{abstract}
The problem of self-consistently coupling kinetic runaway-electron physics to the macroscopic evolution of the plasma is addressed by dividing the electron population into a bulk and a tail. A probabilistic closure is adopted to determine the coupling between the bulk and the tail populations, preserving them both as genuine, non-negative distribution functions. Macroscopic one-fluid equations and the kinetic equation for the runaway-electron population are then derived, now displaying sink and source terms due to transfer of electrons between the bulk and the tail. 
\end{abstract}

\maketitle

\input{intro}
\input{closure}
\input{coupling}

\input{interaction}
\input{summary}

\bibliographystyle{unsrt}
\bibliography{bibfile}    

\end{document}

%% file: intro.tex
\section{Introduction}
In many plasma physics problems of interest it is near impossible to advance the full distribution functions of multiple particle species. To make progress and build intuition, it is thus a common theoretical practice to seek to represent the bulk of the distributions as a Maxwellian parametrized by the associated fluid quantities, and advance the remainder of the distribution function with kinetic principles. Given a source of a small population of highly energetic particles (mono-energetic beam injection, cosmic rays), the kinetic population can be somewhat separately treated, even if the energy carried by the kinetic population is comparable to that carried by the fluid, as the source of the kinetic population is independent of the physics of the bulk population. However, if an imbalance of effective forces generates the kinetic population from the bulk, and the interaction between the particles and the fluid is a key aspect in determining the outcome, then the two populations must be treated together to obtain a fully self consistent model of the evolution. Unfortunately, how to address this type of problem in general remains largely unsolved.  

Among many problems which fit this description, few are more urgent than the modeling of runaway-electron physics in magnetic-confinement fusion experiments. Runaway electrons~\cite{Cohen-1976,Knoepfel-Spong-review-1979}, occurring in tokamaks during plasma-terminating disruptions, are expected to be a serious problem for larger machines such as ITER. The highly-energetic runaway-electron population has been predicted to take over a significant portion of the plasma current during the current-quench phase~\cite{Russo-Campbell:1993} and the concerns for the machine integrity have been amplified after the discovery of the so-called runaway-avalanche effect~\cite{Jayakumar-1993,rosenbluth_putvinski_1997}. The implications of runaway-related issues have been reviewed in detail~\cite{Boozer-2015,Breizman-2017} and multiple studies addressing the runaway formation during disruptions have been carried out, incorporating varying degrees of detail. Some employ approximative analytical results to account for the Dreicer and avalanche mechanisms and estimate the self-consistent current evolution from the combination of Faraday and Amp\`ere laws in a cylindrical plasma column~\cite{Eriksson-et-al:2004,Solis-Loarte-Lehnen-Avalanche-generation-2015,Solis-loarte-lehnen:2017}. Some focus on more detailed modeling of the velocity-space structure of the runaway-electron distribution but consider the plasma as a homogenous wire ignoring all spatial structure~\cite{Stahl:2015,Hesslow:2017}. Finally, there are recent investigations focusing also on runaway-electron formation during impurity dominated thermal quenches~\cite{Aleynikov:2017}, though again neglecting spatial structure. Orbit-following models supplement the palette with tools to estimate effective transport coefficients~\cite{Geri:2015,Konsta-advection-diffusion:2016,Carbajal:2017:orbits} and realistic synchrotron emission signals~\cite{Carbajal:2017:synchrotron,Hoppe:2018}. The given list of references is not comprehensive.

In spite of exhaustive reporting of new results, consensus has not been reached regarding the mitigation strategy of runaway electrons~\cite{ITER_DMS_workshop_report_2017}. One of the most pressing uncertainties lies in the generation and confinement of runaway electrons and the evolution of the macroscopic plasma during the thermal-quench phase: the avalanche amplification of the runaway population during the later current-quench-phase, largely responsible for the conversion of the Ohmic current to relativistic runaway current, is exponentially sensitive to the runaway-electron seed population that emerges during the thermal-quench phase. While it would be of critical importance to evaluate the effect of magnetic flux-surface opening and restoration during the thermal quench on the runaway-electron seed formation and confinement, existing realistic studies evolve marker populations, with no knowledge on how many runaway electrons the markers represent \cite{Izzo_et_al_2011,Sommariva_et_al_2017}. The few self-consistent studies, on the other hand, typically rely on different versions of MHD and model the runaway population as an advected fluid density~\cite{Helander_et_al_2007_resistive_stability,Cai_Fu_2015} traveling at the speed of light, with possible source terms given by analytical estimates for Dreicer and avalanche growth rates~\cite{Matsuyama_et_al_2017}. 

The purpose of this paper is to provide a framework for coupling kinetic runaway-electron dynamics and macroscopic one-fluid evolution self-consistently from first principles. Our intent is to improve upon the existing test-particle models by including the generation process from the bulk to provide an accurate representation of the runaway-electron population, and by accounting for the back response the runaway population has on the macroscopic plasma. To achieve our goals, it is of essence to describe the transfer of particles between the electron bulk and the runaway tail populations as accurately as possible. Unfortunately, neither the popular $\delta f$-approach nor the moment approach, parametrizing the bulk with respect to the fluid quanties and solving for the deviation kinetically, suffice in this context since both have trouble handling long, anisotropic tails that typically characterize runaway-electron populations. As a remedy, we have chosen a recently suggested probabilistic approach\cite{Eero-multiscale} where the coupling between the tail and the bulk is organized in terms of transition probabilities that can be computed deterministically from the stochastic trajectories of test particles~\cite{karney_current_1986,Liu_2016_adjoint,Zhang-DiegoPoP2017-backwards-monte-carlo}. As the new method preserves both the tail and the bulk as genuine, non-negative distribution functions, it, in the end, allows us to write down the one-fluid equations and the runaway-electron kinetic equation in a consistent manner. 

The rest of this paper is organized as follows. In Section II, we briefly review the generic idea of separating the bulk and the tail as presented in Ref~\cite{Eero-multiscale}. Section III is devoted to constructing the fluid-kinetic coupling while Section IV discusses the details necessary for constructing the transition probabilities in the coupling term, closely following the so-called Backward Monte Carlo approach~\cite{Carlsson:2000,Zhang-DiegoPoP2017-backwards-monte-carlo,bormetti_callegaro_livieri_pallavicini_2018} that is more common in option pricing in finance. A discussion regarding how to apply the proposed framework to the runaway-electron-seed formation is given in Section V, while Section VI summarizes our work.

%% file: closure.tex
\section{Probabilistic coupling of bulk and tail}
A fluid-kinetic coupling is not a new idea. It is a relatively common approach to study, e.g., the effects of neutral-beam-driven current or alpha particles on MHD stability~\cite{Belova-denton-chan:1997,Belova-PRL:2015}. In the absence of collisions even elegant variational formulations have been derived \cite{Burby-Tronci:2017}. In case of runaway electrons, the traditional approaches unfortunately are not sufficient. Describing the generation process of runaway electrons accurately requires the presence of collisional processes, and the resulting population is literally dragged out from the bulk forming a long, anisotropic tail. Something different is needed instead. 

Thus we start all the way back from the kinetic equation for species $\alpha$ \begin{align}\label{eq:full-kinetic}
\frac{d f_{\alpha}}{dt}=\sum_{\beta}C_{\alpha\beta}[f_{\alpha},f_{\beta}],
\end{align}
where $d/dt$ refers to a linear phase-space advection operator, such as the Vlasov operator, and $C_{\alpha\beta}$ is a bilinear collision operator between the species $\alpha$ and $\beta$. The collision operator could be the Landau operator representative of the small-angle Coulomb scatterings but it could also include an operator for the close collisions as long as the operator is bilinear in nature. Similarly, the advection operator $d/dt$ could include dissipative forces such as the Landau-Lifshitz expression for radiation damping. Additional source or sink terms could be included if necessary. 

We exploit the linearity of the $d/dt$ operator, the bilinearity of the collision operator, and split the distribution functions according to $f_{\alpha}=f_{\alpha 0}+f_{\alpha 1}$ for each species. Then we introduce a formal split of Eq.\eqref{eq:full-kinetic} according to 
\begin{align}
\frac{d f_{\alpha 0}}{dt}&=\sum_{\beta}C_{\alpha\beta}[f_{\alpha 0},f_{\beta 0}]+\sum_{\beta}C_{\alpha\beta}[f_{\alpha 0},f_{\beta 1}]-I_{\alpha},\\
\frac{d f_{\alpha 1}}{dt}&=\sum_{\beta}C_{\alpha\beta}[f_{\alpha 1},f_{\beta 1}]+\sum_{\beta}C_{\alpha\beta}[f_{\alpha 1},f_{\beta 0}]+I_{\alpha}.
\end{align}
It is straightforward to verify that the sum of the above two equations exactly reproduces Eq.\eqref{eq:full-kinetic}, and thus also any existing conservation laws. For each species, the coupling term $I$, as suggested in Ref\cite{Eero-multiscale}, has the form 
\begin{multline}\label{eq:I}
I(\bm{z},t)=\frac{f_{0}}{\tau}\left(1-\mathbb{E}[\mathbf{1}_{\Omega_0}(\bm{Z}_{t+\tau})|\bm{Z}_t=\bm{z}]\right)\\-\frac{f_{1}}{\tau}\mathbb{E}[\mathbf{1}_{\Omega_0}(\bm{Z}_{t+\tau})|\bm{Z}_t=\bm{z}],
\end{multline}
where $\tau$ is a characteristic time scale, the domain $\Omega_0$ refers to a ``bulk domain'', $\bm{z}$ denotes the phase-space location corresponding to $(\bm{x},\bm{v})$ in case of full particle dynamics, and $\bm{Z}_s$, with $s\in[t,t+\tau]$, denotes the trajectory of an individual particle in the phase-space. The operators $\mathbb{E}$ and $\mathbf{1}$ refer to an expectation value and an indicator function respectively. 

More specifically, $\mathbb{E}[\mathbf{1}_{\Omega_0}(\bm{Z}_{t+\tau})|\bm{Z}_t=\bm{z}]$ is the probability for finding a particle with an initial position $\bm{z}$ at time $t$ within the domain $\Omega_0$ after the time interval $\tau$. Thus, as appearing in the equation for $f_1$, the interaction term adds particles from the bulk at a rate that is proportional to the number of bulk particles available and the probability of a bulk particle leaving the bulk domain. At the same time, it depletes particles from $f_1$ at a rate that is proportional to the number of particles available in the $f_1$ population and the probability of such particles to remain within the bulk domain. As appearing in the equation for $f_0$, the opposite is true of course. 

The details for computing the expectation value $\mathbb{E}[\mathbf{1}_{\Omega_0}(\bm{Z}_{t+\tau})|\bm{Z}_t=\bm{z}]$ are reviewed in Section IV and closely follow the Backward Monte Carlo procedure as presented in Ref~\cite{Zhang-DiegoPoP2017-backwards-monte-carlo}. For now, we focus on formulating the fluid-kinetic coupling given kinetic equations for a bulk and a tail.

%% file: coupling.tex
\section{fluid--kinetic coupling}
The form of the interaction term \eqref{eq:I} guarantees the non-negativity of $f_{0}$ and $f_{1}$ so that both can be interpreted as genuine distribution functions. Thus the kinetic equations are suitable for further reduction. 

Since we are interested in runaway electrons, we shall assume that only the electron distribution function is split into a bulk and a tail according to $f_{\text{e}}=f_{\text{e}0}+f_{\text{e}1}$. Thus only the electron interaction term $I_{\text{e}}$ is needed. Assuming that $C_{\alpha\beta}$ is a physically relevant collision operator with appropriate density, momentum, and energy conservation laws, it is then a simple matter to follow the standard procedure and derive the one-fluid continuity equation
\begin{align}
\partial_t\varrho+\nabla\cdot(\varrho\bm{u})=-\int m_e I_e d\bm{v},
\end{align}
as well as the one-fluid momentum equation
\begin{multline}\label{eq:momentum-equation}
\varrho(\partial_t\bm{u}+\bm{u}\cdot\nabla\bm{u})=-\nabla\cdot\mathbf{p}+\mu_0^{-1}\nabla\times\bm{B}\times\bm{B}\\
-\sum_{\alpha}\bm{F}_{e1,\alpha}-\int m_e(\bm{v}-\bm{u})I_e d\bm{v}\\+en_{e1}(\bm{E}+\bm{v}_{e1}\times\bm{B}).
\end{multline}
In deriving the fluid equations, we have assumed the bulk to be non-relativistic and, to obtain the final form for the momentum equation, the density equation and the assumed conservation properties of the collision operator have been used. 

The new equations differ from the standard one-fluid equations due to the presence of moments of the interaction term, the collisional momentum transfer rate, $\bm{F}_{e1,\alpha}=\int m_e\bm{v}C_{e\alpha}[f_{e1},f_{\alpha}] d\bm{v}$, between the bulk populations and the electron tail, and the term $en_{e1}(\bm{E}+\bm{v}_{e1}\times\bm{B})$ which arises from accounting for the tail population charge and current in the quasineutrality condition and the Amp\`ere's Law. Note that the pressure $\mathbf{p}$ is the fluid pressure, not containing contributions from the runaway tail, and thus the standard equation of state could be used for closing the momentum equation. In the Ohm's law, one should account for the correct bulk flows, leading to 
\begin{align}
\bm{E}+\bm{u}\times\bm{B}=\eta(\mu_0^{-1}\nabla\times\bm{B}+en_{\text{e}1}\bm{v}_{\text{e}1}),
\end{align}
with $\eta$ the Spitzer resistivity. If needed, one could similarly derive an equation for temperature.

Since only the electron tail is treated kinetically, the fluid equations are supplemented with the following equation for $f_{e1}$
\begin{align}\label{eq:runaway-kinetic}
\frac{df_{e1}}{dt}=\sum_{\alpha}C_{e\alpha}[f_{e1},f_{\alpha}]+I_e.
\end{align}
If the runaway electron density remains small compared to the bulk electron density, as it typically does, the nonlinear collision term $C_{\text{ee}}[f_{\text{e}1},f_{\text{e}1}]$ is inferior to the linear collision term $C_{\text{ee}}[f_{\text{e}1},f_{\text{e}0}]$ involving the bulk electrons and can be dropped as self-collisions by definition do not generate momentum or energy. 

Finally, we note that while the collisional momentum transfer rate between the runaway electrons and the bulk plasma, i.e., $\sum_{\alpha}\bm{F}_{e1,\alpha}$, may be expected to be small, absence of it would break the momentum conservation in the system as the balancing term is kept in the kinetic equation for runaways due to its essential role in providing collisional drag and limiting the runaway process. However, the momentum conservation is independent of the shape of the distribution functions. Thus it would be very useful to approximate the fluid species, e.g., as local Maxwellians for the purpose of evaluating the collision operator in \eqref{eq:runaway-kinetic} and the momentum transfer rate in \eqref{eq:momentum-equation}. This assumption could be extended also to the expression of $f_{\text{e}0}$ in the electron interaction term $I_{\text{e}}$. Additionally, during a thermal quench, the expression for $f_{\text{e}0}$ in the interaction term could be adapted to include also the hot-tail effects~\cite{Smith:2008-hot-tail} to better account for the thermal collapse of the plasma. In the test-particle collision operator, these effects most likely would be negligible.

%% file: interaction.tex
\section{Interaction term}
The interaction term requires us to define the time-scale $\tau$, the domain $\Omega_0$, and the probability of finding a test particle within $\Omega_0$ after the time $\tau$. Considering that the fluid equations are eventually discretized in time, the time scale $\tau$ could be chosen to be the time step used in integrating the fluid equations. This way the plasma quantities needed in following test-particle trajectories during $s\in[t,t+\tau]$ for determining the expectation value $\mathbb{E}[\mathbf{1}_{\Omega_0}(\bm{Z}_{t+\tau})|\bm{Z}_t=\bm{z}]$ can be considered constant in time. This assumption turns out to be quite useful. 

\subsection{The general recipe}
The most straightforward way of determining the expectation value would be to perform forward Monte Carlo simulation of multiple test-particle trajectories. As the individual test-particles do not interact with each other, the procedure could in principle be parallelized in an efficient manner, without compromise on the accuracy of the characteristics. In a framework where the fluid evolution is solved alongside the kinetic distribution function, it can, however, be a nontrivial task to perform the parallelization efficiently in practice: the macroscopic data is typically domain decomposed for an efficient parallel solve of the fluid equations, and following the particle characteristics over different domains requires efficient communication between the compute nodes as well as good load-balancing to avoid idling of the compute nodes. And as the kinetic equation for runaways could be solved also with grid-based methods, we wish to discuss efficient methods to compute the necessary expectation value. The proposed approach is based on the connection between time-independent Fokker-Planck equations and the Feynman-Kac formula for expectations of stochastic processes\cite{karatzas2012}. Often referred to as the Backward Monte Carlo method, it is commonly encountered in option pricing in finance\cite{bormetti_callegaro_livieri_pallavicini_2018} but has been recently applied to determine also the electron runaway probability in plasmas\cite{Zhang-DiegoPoP2017-backwards-monte-carlo}. Also other methods for computing the expectation value, such as the adjoint formulation\cite{karney_current_1986,Liu_2016_adjoint}, could be adopted. For a detailed discussion, see Refs\cite{Zhang-DiegoPoP2017-backwards-monte-carlo,Eero-multiscale}.

In determining the transition probability, the test-particle motion is assumed to obey a stochastic differential equation, or a so-called Langevin equation. This model can capture the deterministic motion due to electromagnetic fields, stochastization corresponding to a Fokker-Planck collision operator and quasilinear diffusion effects, as well as radiation reaction force that can become important at high energies. The test-particle motion modeled by a stochastic differential equation does not capture the close collisions as they cannot be expressed in a Fokker-Planck form. Nevertheless, since the role of the interaction term is only to relabel particles between the bulk and the tail populations while explicitly not changing their phase-space position, the neglect of close collisions in determining the relabeling is most likely acceptable. In the end, once the particles have been relabeled from the fluid to the kinetic population, the kinetic equation used for advancing the particle distribution in phase-space does contain the close-collision operator. 

We consider here the full particle motion as an example for mapping the test-particle characteristics. Also guiding-center motion could be exploited whenever applicable. The choice for the time scale $\tau$ to correspond to the fluid time-step allows us to treat the background plasma independent of time during the interval $s\in[t,t+\tau]$ and the stochastic differential equation for the test-particle characteristics $\bm{Z}_s=(\bm{X}_s,\bm{P}_s)$ in It\^o convention is then given by
\begin{align}
d\bm{X}_s&=\bm{V}_sds,\\
d\bm{P}_s&=\bm{\mu}(\bm{Z}_s,t)ds+\bm{\sigma}(\bm{Z}_s,t)\cdot d\bm{W}_s,
\end{align}
where $\bm{V}_s=\bm{P}_s/(m\gamma_s)$ with $\gamma_s=\sqrt{1+|\bm{P}_s|^2/m^2c^2}$ the relativistic factor, and the coefficients $\bm{\mu}$ and $\bm{\sigma}$ are evaluated with respect to time $t$ but at the particle position $\bm{Z}_s$, and $\bm{W}_s$ is a standard vector-valued Wiener process. 
If the stochastic contribution to the test particle motion would satisfy $||\bm{\sigma}||^2\tau\leq 1$, one could estimate $\bm{Z}_{t+\tau}$ by discretizing the stochastic differential with, e.g., the Euler-Maruyama scheme using only one time step, substituting the result to the indicator function, and then deterministically evaluate the expectation value as an average over the distribution of of the standard Normal random variable that would appear in the discretization of the stochastic differential equation.

In general, the condition $||\bm{\sigma}||^2\tau\leq 1$ might not be true. To nevertheless avoid the forward Monte Carlo simulation, it is possible to exploit the Feynman-Kac formula on the basis of the assumption that the coefficients $\bm{\mu}$ and $\bm{\sigma}$ are assumed constant in time during $s\in[t,t+\tau]$. As explained in Refs.\cite{Eero-multiscale,Zhang-DiegoPoP2017-backwards-monte-carlo}, one first defines a quantity
\begin{align}
\Phi(\bm{z};s)=\mathbb{E}[\mathbf{1}_{\Omega_0}(\bm{Z}_{t+\tau})|\bm{Z}_s=\bm{z}],
\end{align}
so that $\Phi(\bm{z};t)=\mathbb{E}[\mathbf{1}_{\Omega_0}(\bm{Z}_{t+\tau})|\bm{Z}_t=\bm{z}]$ and $\Phi(\bm{z};t+\tau)=\mathbf{1}_{\Omega_0}(\bm{z})$. Then one introduces the partition of the interval $s\in[t,t+\tau]$ according to $\{t=s_0,s_1,...,s_N=t+\tau\}$ and, for every point $\bm{z}=(\bm{x},\bm{p})$ of interest, recursively computes 
\begin{align}\label{eq:probability}
\Phi(\bm{z},s_{n-1})=\int_{\mathbb{R}^3}\Phi(\bm{Z}_{\Delta s}(\bm{w}),s_n)\frac{\exp[-\bm{w}^2/2]}{(2\pi)^{3/2}}d\bm{w},
\end{align}
where $\Delta s=s_{n}-s_{n-1}$ and the test particle position
 $\bm{Z}_{\Delta s}(\bm{w})=(\bm{X}_{\Delta s},\bm{P}_{\Delta s}(\bm{w}))$ after the interval $\Delta s$ is estimated from 
\begin{align}
\bm{X}_{\Delta s}&=\bm{x}+\int_{0}^{\Delta s}\frac{d \bm{x}}{ds}ds\\
\widetilde{\bm{P}}_{\Delta s}&=\bm{p}+\int_{0}^{\Delta s}\frac{d \bm{v}}{ds}ds,\\
\bm{P}_{\Delta s}&=\widetilde{\bm{P}}_{\Delta s}+\bm{\sigma}\left(\bm{X}_{\Delta s},\widetilde{\bm{P}}_{\Delta s},t\right)\cdot\bm{w}\sqrt{\Delta s},
\end{align}
with the deterministic trajectory obeying the ordinary differential equations $d\bm{x}/ds=\bm{v}$ and $d\bm{p}/ds=\bm{\mu}$. The integral \eqref{eq:probability} can be computed efficiently to high accuracy using Gauss-Hermite quadrature rules as demonstrated previously~\cite{Zhang-DiegoPoP2017-backwards-monte-carlo}. One should note that for each point $\bm{z}$ the values $(\bm{X}_{\Delta s},\widetilde{\bm{P}}_{\Delta s})$ are needed only once regardless of how many intervals $\Delta s$ are used for the recursive computation. This is expected to offer a significant computational benefit compared to forward Monte Carlo simulation.

As a final note, the domain $\Omega_0$ should be chosen so that the resulting population $f_{e1}$ would be characteristic of runaway electrons. In solving the kinetic equation, we do not wish to waste resources on particles that would not become runaways with reasonable probability. A sufficient choice would thus limit the bulk domain boundary, e.g., to some multiple of the so-called critical velocity. 

\subsection{Simplifications}
Under specific conditions, the evaluation of the expectation value can be simplified somewhat further, making it computationally more appealing. As our measure for whether a particle is to be relabeled into a kinetic population depends on it's likelihood to end up beyond some boundary in velocity, we are ultimately interested in what happens to the single test particle's energy. If we are willing to accept the assumption that electrons would remain approximately fixated to a field line, and that the macroscopic fluid quantities would remain approximately constant along the field lines, then, in computing the expectation value, we could simplify the stochastic differential equation significantly. To lowest order, we could consider, e.g., only the motion in 2D velocity space space that is aligned with respect to the local direction of the magnetic field. While this assumption is common in reduced modeling of runaway electron distribution functions, for us it serves only the purpose of estimating the transition probabilities for relabeling fluid particles to the kinetically treated population, which can then be simulated with accurate orbit following techniques. 

To provide an example, we consider the test particle dynamics to be given by a  relativistic Fokker-Planck equation including the acceleration due to an electric field parallel to the magnetic field, scattering and drag from the small-angle Coulomb collisions, and the synchrotron radiation reaction force. Details of this model can be found in Ref.\cite{Zhang-DiegoPoP2017-backwards-monte-carlo}. The stochastic equations for the electron momentum $p$ and pitch angle cosine $\xi=\bm{p}\cdot\bm{B}/(pB)$ are given by
\begin{equation}\label{e1}
\begin{aligned}
d p_s & =\mu_p(p_s, \xi_s)\, ds, \\
d \xi_s & = \mu_{\xi}(p_s, \xi_s)\, ds + \sigma_{\xi} (p_s,\xi_s)\, dW_s,
\end{aligned}
\end{equation}
where the drift coefficients $\mu_p, \mu_\xi$ and the diffusion coefficient $\sigma_\xi$ are given by
\begin{equation}\label{e2}
\begin{aligned}
& \mu_p = E \xi - \frac{\gamma p}{\tau_r}  \left( 1- \xi^2 \right)-\frac{1+p^2}{p^2},\\
& \mu_{\xi} =  \frac{E \left(1-\xi^2\right)}{p} + \frac{\xi \left(1-\xi^2\right)}{\tau_r \gamma} -\xi \nu_c,\\
&  \sigma_{\xi} =  \sqrt{ \nu_c \left(1-\xi^2\right)}, \,
\end{aligned}
\end{equation}
with $\nu_c = \left(Z+1\right) \sqrt{1+p^2}/p^3$. Here normalized units have been used. The momentum is normalized to $m_ec$, $\gamma=\sqrt{1+p^2}$ is the relativistic factor, $Z$ is the effective charge, the electric field has been normalized to the Connor-Hastie critical electric field,
$E_c=n_e e^3 {\rm ln} \Lambda/(4 \pi \varepsilon_0^2 m_e c^2)$, where $\Lambda$ is the Coulomb logarithm, 
and time has been normalized using the relativistic collision time scale, $\tau_{c}=m_e c/(E_c e)$. The parameter
${\tau_r}=6 \pi \varepsilon_0 m_e^3 c^3/(e^4 B^2\tau_c)$
is the normalized synchrotron radiation time scale with $\varepsilon_0$ the vacuum permittivity.

\begin{figure}[!h]
\includegraphics[width=\columnwidth]{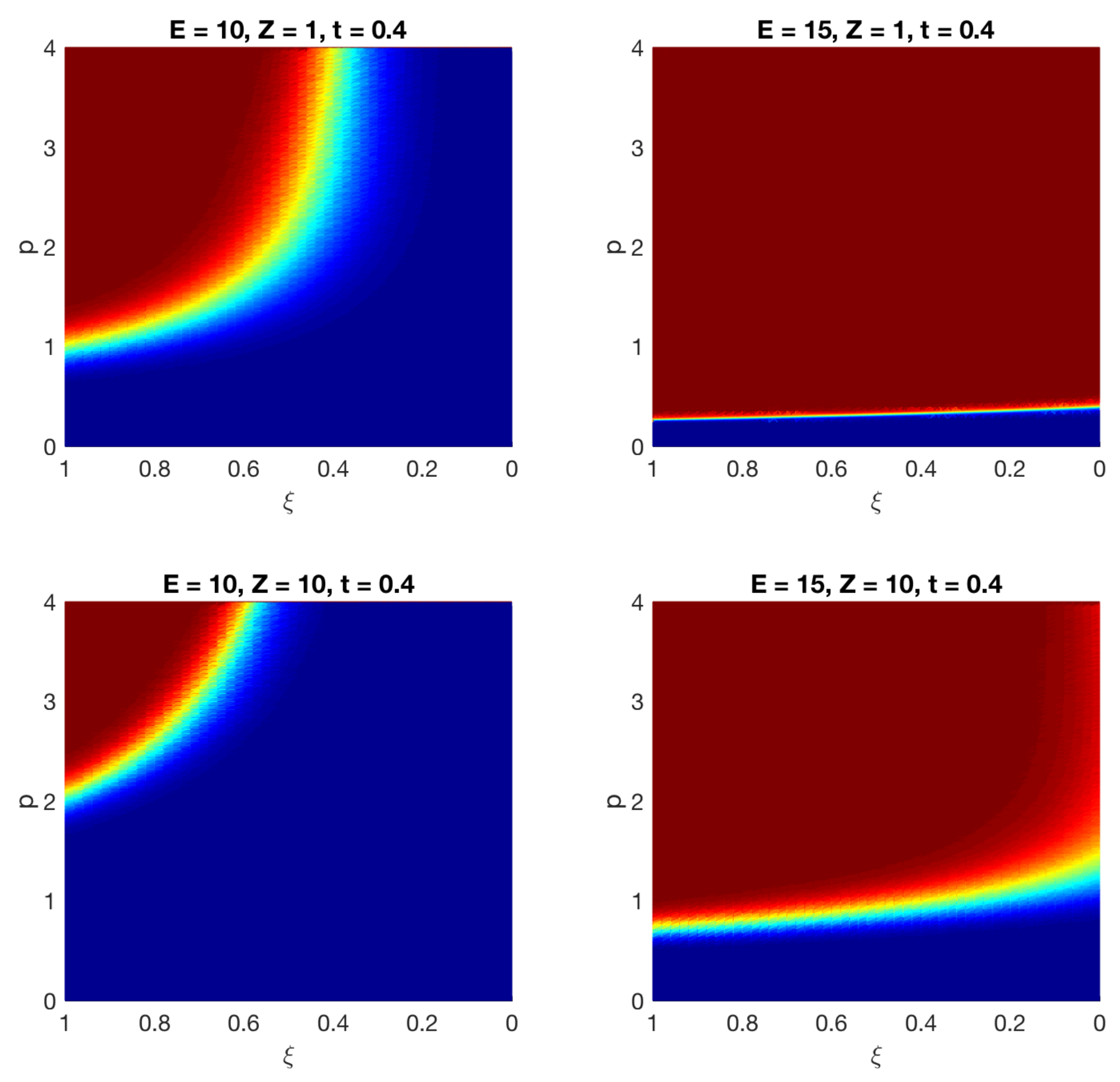}
\caption{\label{fig:example} Computation of $1-\Phi$, i.e., the runaway probability, using the simplified 2D momentum space model for particle characteristics. For illustration purposes we fixed $\tau_r=1$, $t=0.4$ and changed $Z$ and $E$. The boundary in this case was assumed to be at $p=4$. As expected, for a fixed valued of $Z$, the runaway region increases with $E$. On the other hand, for fixed $E$, the runaway region decreases with increasing $Z$. The red color corresponds to $\Phi=0$ (particles become runaways with certainty) while blue corresponds to $\Phi=1$ (particles remain in the bulk). The implementation details regarding the evaluation of the Gaussian-weighted integrals can be found in Ref.\cite{Zhang-DiegoPoP2017-backwards-monte-carlo}.}
\end{figure}

The parameters of the model are, $E$, $\tau_r$ and $Z$, and would be determined in terms of the macroscopic background quantities. As such, this model for determining the transition probabilities is entirely local in the configuration space and could be implemented efficiently with respect to any domain decomposition the fluid solver requires. As an illustration, Fig.~\ref{fig:example} shows $1-\Phi$ with respect to different values for the parameters $E$, $\tau_r$ and $Z$, demonstrating the importance of pitch-angle physics in capturing the runaway transition probability. Once a fast computational tool for evaluating $\Phi$ exists, it can be used in conjunction with the kinetic equation for runaways and the fluid equations to accurately and efficiently capture the generation process of runaway electrons in complicated magnetic fields that occur during a thermal quench.

%% file: summary.tex
\section{Discussion}
To make also practical progress in the runaway electron issues, we would next have to apply the presented ideas. The lowest hanging fruit would be to address the number of seed runaway electrons during a thermal quench, before a significant runaway density or current forms. In this case, it would most likely be sufficient to follow the evolution of $f_{\text{e}1}$ while ignoring the knock-on collisions and the effects $f_{\text{e}1}$ would pose on the fluid. One could further ignore the transfer of $f_{\text{e}1}$ particles back to the bulk so that the interaction term could be approximated as
\begin{align}
I_e=\frac{f_{\text{e}0}}{\tau}\left(1-\mathbb{E}[\mathbf{1}_{\Omega_{\text{e}0}}(\bm{Z}_{t+\tau})|\bm{Z}_t=\bm{z}]\right).
\end{align}
In this case, the kinetic equation for the runaway distribution becomes linear, containing only the advection in phase-space, test-particle collisions with fluid species, and the interaction term that now depends only on the fluid quantities. One way to solve the evolution of $f_{\text{e}1}$ in this setting would involve populating the phase-space with marker particles with initially zero weights, following their orbits, applying Monte Carlo coulomb collisions, and increasing the marker weights according to the interaction term. This would result in an algorithm that would be straightforward to include to append into existing MHD codes that support marker-particle following with as little modification as possible.

\section{Summary}
The problem of self-consistently coupling kinetic runaway-electron physics to the macroscopic evolution of the plasma was addressed by dividing the electron population into a tail and a bulk. A probabilistic closure was adopted to determine the coupling between the bulk and the tail populations, preserving them both as genuine, non-negative distribution functions. Macroscopic one-fluid equations and the kinetic equation for the runaway-electron population were then derived, displaying sink and source terms due to transfer of electrons between the bulk and the tail. The details needed in computing the interaction term were provided together with a simple numerical example. In near future, the model could serve to determine how many runaway electrons are generated during the plasma terminating disruptions, to accurately assess the threat the runaway electrons project on ITER operation.

\begin{acknowledgments}
The Authors are grateful for the encouragement and support from Professor Amitava Bhattacharjee and the SCREAM collaboration. EH, CL, and DB acknowledge the support from the the U.S. Department of Energy Contract No. DE-AC02-09-CH11466 and DE-SC0016268. DdCN and GZ acknowledge support from the U.S. Department of 
Energy at Oak Ridge National Laboratory, managed by UT-Battelle, LLC, 
for the U.S. Department of Energy under contract DE-AC05-00OR22725.
The views and opinions expressed herein do not necessarily reflect those of the U.S. Department of Energy.
\end{acknowledgments}